\documentclass[twocolumn,floatfix,showpacs,prd,aps,tightenlines,superscriptaddress]{revtex4}
\usepackage{graphicx}
\usepackage{amsmath}
\usepackage{amssymb}
\usepackage{mathrsfs}

\usepackage{dcolumn}
\usepackage{bm}

\newcommand{\bea}{\begin{eqnarray}}
\newcommand{\eea}{\end{eqnarray}}
\newcommand{\beq}{\begin{equation}}
\newcommand{\eeq}{\end{equation}}
\newcommand{\KMS}{{\rm km\ s^{-1}}}

\begin{document}

\title{Extra-Large Remnant Recoil Velocities and Spins from
Near-Extremal-Bowen-York-Spin Black-Hole Binaries}

\author{Sergio Dain}
\affiliation{Facultad de Matem\'atica,
  Astronom\'{\i}a y F\'{i}sica, Universidad Nacional de C\'ordoba,
  Ciudad Universitaria (5000) C\'ordoba, Argentina.  }

\affiliation{Max Planck Institute for Gravitational Physics (Albert
  Einstein Institute) Am M\"uhlenberg 1 D-14476 Potsdam Germany.}

\author{Carlos O. Lousto}
\affiliation{Center for Computational Relativity and Gravitation,
School of Mathematical Sciences,
Rochester Institute of Technology, 78 Lomb Memorial Drive, Rochester,
 New York 14623}

\author{Yosef Zlochower}
\affiliation{Center for Computational Relativity and Gravitation,
School of Mathematical Sciences,
Rochester Institute of Technology, 78 Lomb Memorial Drive, Rochester,
 New York 14623}

\date{\today}

\begin{abstract}
We evolve equal-mass, equal-spin black-hole binaries with  specific
spins of $a/m_{H} \sim 0.925$, the highest spins simulated thus far
 and nearly the largest possible for
Bowen-York black holes,  in a set of configurations with the spins
counter-aligned and pointing in the orbital plane, which  maximizes
the recoil velocities of the merger remnant, as well as a
configuration where the two spins point in the same direction as the
orbital angular momentum, which maximizes the orbital hang-up effect
and remnant spin.  The coordinate radii of the individual apparent
horizons in these cases are very small and the simulations  require
very high central resolutions ($h\sim M/320$). We find that these
highly spinning holes reach a maximum recoil velocity of $\sim 3300\
\KMS$ (the largest simulated so far) and, for the hangup
configuration,
a remnant spin of $a/m_{H} \sim 0.922$.  These results are consistent
with our previous predictions for the maximum recoil velocity of
$\sim4000\ \KMS$ and
remnant spin; the latter reinforcing the prediction that cosmic
censorship is not violated by merging highly-spinning black-hole
binaries. We also numerically solve the initial data for, and evolve, a
single maximal-Bowen-York-spin black hole, and confirm that the 3-metric has
an ${\cal O}(r^{-2})$ singularity at the puncture, rather than the
usual ${\cal O}(r^{-4})$ singularity seen for non-maximal spins.
\end{abstract}

 \maketitle

\section{Introduction}\label{sec:introduction}

Highly spinning black holes play an important role in astrophysics,
from powering active galactic nuclei (AGN), to $\gamma$-ray bursts
(GRB) and quasars.  While the (indirect) observational evidence for
the existence of black holes, stellar mass or supermassive, is
overwhelming, the actual observational evidence for black-hole spin is
scarce.  There have been attempts to measure the central black hole
spin in AGN~\cite{Reynolds:2004qk, Lodato:2006kv}, Seyfert
galaxies~\cite{Wilms:2001cm}, and quasars~\cite{Wang:2006bz}.  The
x-ray spectra of accretion disks around stellar mass black holes can
also provide information about their spins~\cite{Miniutti:2003wp,
Zhang:1997dy, Reynolds:2007rx}.

The recent dramatic breakthroughs in the numerical techniques to
evolve black-hole-binary spacetimes~\cite{Pretorius:2005gq,
Campanelli:2005dd, Baker:2005vv} has led to rapid advancements in our
understanding of black-hole physics.  Notable among these advancements
are developments in mathematical relativity, including systems of PDEs
and gauge choices~\cite{Lindblom:2005qh,Gundlach:2006tw,
vanMeter:2006vi}, the exploration of the cosmic
censorship~\cite{Campanelli:2006uy, Campanelli:2006fg,
Campanelli:2006fy, Rezzolla:2007rd, Sperhake:2007gu},  and the
application of isolated horizon formulae~\cite{Ashtekar:2000hw,
Dreyer:2002mx, Schnetter:2006yt, Campanelli:2006fg, Campanelli:2006fy,
Krishnan:2007pu}.  These breakthroughs have also influenced the
development of data analysis techniques through the matching of
post-Newtonian to fully-numerical waveforms~\cite{Pan:2007nw,
Boyle:2007ft, Hannam:2007ik,Ajith:2007kx}. In particular, the moving punctures
approach proved to work in a wider realm than was originally
thought. Notably, it has been successfully applied to many-black-hole
spacetimes~\cite{Campanelli:2007ea, Lousto:2007rj}, and to
black-hole--neutron star evolutions~\cite{Shibata:2006bs,
Shibata:2006ks, Shibata:2007zm, Etienne:2007jg}.  Similarly, the
recent discovery of very large merger recoil
kicks~\cite{Campanelli:2007ew, Gonzalez:2007hi, Campanelli:2007cga,
Herrmann:2007ac, Koppitz:2007ev, Herrmann:2006ks, Baker:2006vn,
Gonzalez:2006md} has had a great impact in the astrophysical
community, with several groups now seeking for observational traces of
such high speed holes as the byproduct of galaxy
collisions~\cite{Bogdanovic:2007hp, Loeb:2007wz, Bonning:2007vt,
HolleyBockelmann:2007eh}.

The
first study of generic black-hole-binary configurations (i.e.\
binaries with unequal component masses and spins, and spins not
aligned with each other or the orbital angular momentum) was described
in Ref.~\cite{Campanelli:2007ew}, and, based on the results of that
study, a semi-empirical formula to estimate the recoil velocities of
the remnant black holes was proposed, finding recent confirmation
in~\cite{Campanelli:2007cga, Herrmann:2007ex, Brugmann:2007zj}.
The spin contributions to the
recoil velocity are generally larger than those due to the unequal
masses, and, in particular, the spin component in the orbital plane
has the largest effect~\cite{Campanelli:2007ew}, leading to a maximum
recoil velocity of about $3500 - 4000\ \KMS$~\cite{Campanelli:2007cga}.
The recoil velocities acquired by the remnant of the merger of
black-hole binaries has many interesting astrophysical
consequences~\cite{Campanelli:2007ew}, particularly since spinning
black holes can accelerate the merged hole high enough to eject the remnant
from the host galaxy. Recently a quasar, displaying blue shifted
emission lines by $2650\,km/s$ with respect to its host galaxy,
has been observed \cite{Komossa:2008qd}.

In all the above simulations, the evolution was started using
conformally flat initial data. This choice has the advantage of being
easy to implement, with the (apparently) minor drawback of introducing
a short, non-physical burst of radiation at the start of the
simulation. Apart from this initial burst, there appears to be no
unphysical behavior associated with conformally flat initial data, and
this choice remains popular (See Refs.~\cite{Tichy02, Kelly:2007uc}
for more astrophysically realistic initial data using post-Newtonian
information).  The simplest initial data, Bowen-York (BY), gives the
extrinsic curvature, $K_{ij}$ analytically by assuming that it is
transverse and traceless (See~\cite{Dain:2002ee} for an alternative prescription).
 A particularly interesting feature of the
conformally flat ansatz for the 3-metric is that these data cannot
model maximally rotating Kerr black holes (such holes are not
conformally flat in any smooth slice), but have a limiting
spin~\cite{Cook90a,Dain:2002ee} of $S/M_{ADM}^2\approx0.928$ for BY data and
$S/M_{ADM}^2\approx0.932$ for conformally Kerr extrinsic
curvature~\cite{Dain:2002ee}. Here $S$ denotes the spin of the black
hole and $M_{ADM}$ the total ADM mass.

The spin of the merger remnant is similarly important both because
high-spin black holes are more efficient at converting accreting matter into
radiation than lower-spin holes, and because of open questions
regarding cosmic censorship. This issues have already been studied in
the `Lazarus' approach to numerical evolutions~\cite{Baker:2004wv} and
in early evolutions using the `moving punctures'
approach~\cite{Campanelli:2006uy, Campanelli:2006fg,
Campanelli:2006fy}.  Recently the issue has been revisited in the
context of unequal mass holes~\cite{Rezzolla:2007xa, Rezzolla:2007rd,
Sperhake:2007gu, Buonanno:2007sv, Rezzolla:2007rz},
and highly-elliptical, equal-mass
binaries~\cite{Washik:2008jr,Hinder:2007qu}.  In the current
work we show that, for the maximum possible spin for Bowen-York black
holes, the merger remnant will always have sub-maximal spins.

This paper is organized as follows: in Sec.~\ref{sec:BYID} we
describe the Bowen-York initial data for a single spinning black hole
and how one can obtain the maximum possible spin for a Bowen-York
black hole, in Sec.~\ref{sec:ID} we describe how we obtain initial
data for black-hole binaries with nearly maximal BY spin, in
Sec.~\ref{sec:Techniques} we describe the numerical techniques used to
evolve these binaries, in Sec.~\ref{sec:results} we give the results
and analysis from the numerical evolutions, and in
Sec.~\ref{sec:conclusion} present our conclusions.

\section{Initial Data}

An \emph{initial data set} for the Einstein vacuum equations is given by a
triple  $({\cal M},   \gamma_{ij},   K_{ij})$  where ${\cal M}$
is a connected 3-dimensional manifold, $  \gamma_{ij} $ a (positive
definite) Riemannian metric, and $  K_{ij}$ a symmetric tensor
field on 
${\cal M}$, such that the constraint
equations
\begin{align}
 \label{const1}
   D_j   K^{ij} -  D^i   K=J^i,\\
 \label{const2}
   R -  K_{ij}   K^{ij}+  K^2=2\mu ,
\end{align}
are satisfied on ${\cal M}$. Where $ {D}$ and $  R$ are the
Levi-Civita connection and the Ricci scalar associated with
$ {\gamma}_{ij}$, and $K =   K_{ij}   \gamma^{ij}$. The vector $J^i$ and the
scalar function $\mu$ are determined by the stress energy tensor of
the sources which describes the matter content of the spacetime.  In
these equations the indices are moved with the metric $  \gamma_{ij}$
and its inverse $  \gamma^{ij}$.

\subsection{Maximum Spin Bowen-York initial data}\label{sec:BYID}

The Bowen-York family of initial sets~\cite{Bowen80} represents a
relevant class of data suitable for numerical simulations of
black-hole binaries.  They are constructed using the conformal method for
solving the constraint equations~\eqref{const1}--\eqref{const2} (for a
recent review on this method see~\cite{Bartnik04b} and references
therein).  Let us consider one member of this family, namely the
spinning Bowen-York data. These data describe a (non-stationary) black
hole with intrinsic angular momentum. In order to construct the data
we prescribe $\tilde K^{ij}$, a symmetric, trace free and divergence free tensor
with respect to the flat metric $\delta_{ij}$
\begin{equation}
  \label{eq:51}
  \tilde K^{ij} =\frac{6}{r^3}n^{(i}  \epsilon^{j)kl} S_k n_l,  
\end{equation}
where $r$ is the spherical radius, $n^i$ the corresponding radial
unit normal vector, $\epsilon^{ijk}$ the flat volume
element  and $S_k$ an arbitrary  constant vector which will give the
total spin of the data. In this equation the
indices are moved with the flat metric $\delta_{ij}$. 
The data are given by
\begin{equation}
  \label{eq:4}
  \gamma_{ij} =
\varphi^4  \delta_{ij} \quad {K}^{ij} = \varphi^{-10}{\tilde K}^{ij}, 
\end{equation}
where the conformal factor satisfies the following equation (which is
the conformal version of the Hamiltonian constraint~\eqref{const2})
\begin{equation}
  \label{eq:86}
  \Delta \varphi  =- \frac{18 S^2 \sin^2\theta}{8r^6\varphi^7}.
\end{equation}
where $S^2=S_iS_j\delta^{ij}$ and $\Delta$ is the flat Laplacian.

For any solution $\varphi$ of Eq.~\eqref{eq:86} the metric
$\gamma_{ij}$ and $K_{ij}$ given by~\eqref{eq:4} define a solution of the
vacuum (i.e. $J^i=\mu=0$) constraint
 equations~\eqref{const1}--\eqref{const2}.  In order to find a unique and
physically relevant solution of \eqref{eq:86} we need to impose
appropriate boundary conditions for $\varphi$. This is essentially
equivalent to prescribe the manifold ${\cal M}$ of the initial data. For
example, Eq.~\eqref{eq:86} is singular at $r=0$, it follows that
the solution $\varphi$ can not be regular at $r=0$ and hence the
origin can not be in the manifold ${\cal M}$. That is $\mathbb{R}^3$ is not
allowed as manifold in this class of initial data. In the present case
the manifold will be ${\cal M}=\mathbb{R}^3\setminus \{0\}$, the origin $r=0$
represents another end of the initial data.

Boundary conditions for black holes are prescribed as follows.  Let
$m^p>0$ be an arbitrary number. Define the function $u$ by
\begin{equation}
  \label{eq:85}
  \varphi=1+\frac{m^p}{2r}+u.
\end{equation}
Using~\eqref{eq:86} we obtain an equation for $u$ outside the origin 
\begin{equation}
  \label{eq:2}
  \Delta u =- \frac{18r S^2 \sin^2\theta}{8(r+\frac{m^p}{2}+ru)^7}.
\end{equation}
If $u$ is positive (this will be the case as a consequence of the
maximum principle) then the denominator of Eq.~\eqref{eq:2} never
vanishes and hence this equation is regular at the origin. 
The idea is to impose this equation also at the origin, that is, we
want to solve~\eqref{eq:2} in $\mathbb{R}^3$ with a boundary condition
at infinity 
\begin{equation}
  \label{eq:11}
  \lim_{r\rightarrow
    \infty} u =0.
\end{equation}
It is well known that for each $m^p>0$ there exists a unique solution
$u$ of~\eqref{eq:2} which satisfies the boundary condition~\eqref{eq:11}.
 The solution is positive and from standard elliptic
theory it follows that it is smooth outside the origin and it is $C^2$
at the origin (this loss of differentiability is due to the presence
of the function $r$ on the right hand side of Eq.~\eqref{eq:2}
which is not smooth at the origin).  This is what in the
numerical relativity is called the puncture method~\cite{Brandt97b},
the parameter $m^p$ is called the mass parameter of the puncture. 
By Eq.~\eqref{eq:85}, the singular part of $\varphi$ at $r=0$ is
${\cal O}(1/r)$, this implies that the physical fields $\gamma_{ij},K_{ij}$
are asymptotically flat at the end $r=0$.

The physical parameters of the data are given by $S$ which represent
the angular momentum of the data and the total ADM mass $M_{ADM}$, which is
given by the following formula
\begin{equation}
  \label{eq:5}
  M_{ADM}=m^p+m_u,
\end{equation}
where $m_u$ is the coefficient ${\cal O}(1/r)$ of $u$, that is
\begin{equation}
  \label{eq:12}
  u=\frac{m_u}{2r}+{\cal O}(1/r^2)
\end{equation}
as $r\to \infty$. Note that in order to calculate  the total mass $M_{ADM}$ we need to
solve the non-linear equation~\eqref{eq:2}

The solution $u$ depends on the coordinates $x$ and the two parameters
$m^p$ and $S$.  However, there exists a scale invariance for equation
\eqref{eq:2}.  Namely, if we have a solution $u(m^p,S, x)$, for
parameters $S$ and $m^p$, then the rescaled function $u(\lambda m^p,
\lambda^2 S, \lambda x)$, where $\lambda$ is an arbitrary positive
number, is also a solution.  This means that the solution depends non
trivially only on one parameter. We chose to fix $S$ and vary $m^p$.
Note that the following quotient is scale invariant
\begin{equation}
  \label{eq:13}
\epsilon_S=  \frac{S}{M_{ADM}^2}. 
\end{equation}
For a Kerr black hole we have $\epsilon_S\leq 1$ and $\epsilon_S=1$
implies that the black hole is extreme.  For general axially symmetric
vacuum black holes (which in particular includes the spinning
Bowen-York data) we also have $\epsilon_S\leq
1$~\cite{Dain05e}\cite{Dain06c} and $\epsilon_S=1$ if and only if the
data are slices of extreme Kerr black hole. Since the Bowen-York data
are not slices of the extreme Kerr black hole it follows that $\epsilon_S <1$
for this family.  What is the maximum value for $\epsilon_S$ in this
family? This question was explored numerically
in~\cite{Choptuik86,Cook90a,Dain:2002ee}. In these references it was observed
that in the limit $S\to \infty$ (for fixed $m^p$) the ratio
$\epsilon_S$ reach an asymptotic maximum value. Because of
the scaled invariance mentioned above this limit is equivalent to
$m^p\to 0$, with $S$ fixed.  What was not clear at all is that in fact
in the limit we get a well behaved solution of the constraint
equation.  This is precisely the question we want to answer here. That
is, we want to give numerical evidence that the limit
\begin{equation}
  \label{eq:1}
  u(S,x)= \lim_{m^p\to 0} u(m^p,S,x)
\end{equation}
exists and defines a solution of the constraint equations. An
analytical proof of this is detailed in~\cite{Dain:2008yu}.
We will call this new solution the extreme Bowen-York spinning data
because it has the maximum amount of angular momentum per mass in this
family~\footnote{An alternative method for generating highly-spinning
binaries, which was released as a preprint a few months after this paper was
submitted, is given in~\cite{Lovelace:2008tw}}.

In the limit $m^p\to 0$ the difference between $\varphi$ and $u$ is just
a constant and Eq.~\eqref{eq:2} become singular at the origin, hence
 the limit solution $u$ can not be regular at the origin.  If
we assume that $u={\cal O}(r^\alpha)$ at $r=0$ for some real number $\alpha$,
using Eq.~\eqref{eq:2} we get that $\alpha=-1/2$. That is, we
expect the following behavior at the origin
\begin{equation}
  \label{eq:9}
  u={\cal O}\left(\frac{1}{\sqrt{r}}\right).
\end{equation}
This behavior is confirmed by the numerical simulations presented in
the next subsection~\ref{sec:hspinnrtest}.

We have seen that the limit solution $u$ has a different fall off
behavior at the origin than the family $u(m^p)$ for $m^p>0$, in particular
the conformal factor for $m^p>0$ behaves like ${\cal O}(1/r)$ at $r=0$ but in
the limit $m^p\to 0$ it behaves like ${\cal O}(1/\sqrt{r})$. This implies a
change in the fall off behavior off the physical fields at the
end $r=0$. This end will not be asymptotically flat in the extreme
limit.

We illustrate the same phenomena with two important examples. The first
one is the Reissner-Nordstr\"om black hole initial data. In isotropic
coordinates, a canonical slice $t={\rm const}$ for the
Reissner-Nordstr\"om
black hole spacetime with mass $M_{ADM}$ and charge $q$ (for a black hole we
always have $q\leq M_{ADM}$) defines the following initial data set
\begin{equation}
  \label{eq:6}
 \gamma_{ij}=\varphi^4 \delta_{ij}, \quad K^{ij}=0, \quad J^i=0, \quad
 \mu=\frac{q^2}{r^4\varphi^8}, 
\end{equation}
where the conformal factor is given by
\begin{equation}
  \label{eq:83}
  \varphi  = \frac{1}{2r} \sqrt{(q+2r+M_{ADM})(-q+2r+M_{ADM})}.
\end{equation}
The conformal factor satisfies the following equation (analog to
Eq.~\eqref{eq:86})
\begin{equation}
  \label{eq:15}
    \Delta \varphi  =- \frac{q^2}{4r^4\varphi^3}.
\end{equation}
For this solution define the parameter
\begin{equation}
  \label{eq:3}
  m^p=\sqrt{M_{ADM}^2-q^2},
\end{equation}
and the function $u$ by
\begin{equation}
  \label{eq:7}
  u=\varphi-1-\frac{m^p}{2r}. 
\end{equation}
For $m^p>0$, the function $u$ is bounded at $r=0$. The
extreme limit $q\to M_{ADM}$ corresponds to $m^p\to 0$. In this limit
we have
\begin{equation}
  \label{eq:8}
  \varphi=\sqrt{\frac{M_{ADM}}{r}+1 }, \quad u=\varphi-1,
\end{equation}
and hence $\varphi=u={\cal O}(1/\sqrt{r})$. Note that although Eq.~\eqref{eq:15}
 is different from~\eqref{eq:86}, the powers of $r$ and
$\varphi$ on the right hand side are such that if we assume
$u={\cal O}(r^\alpha)$ at $r=0$ we get $\alpha=-1/2$ as in the Bowen-York
spinning black hole.

The second example is given by the Kerr black hole initial data in
quasi-isotropic coordinates. Let $S$ and $M_{ADM}$ be the total angular
momentum and mass of the Kerr spacetime. Define the parameter $m^p$ by 
\begin{equation}
  \label{eq:10}
  m^p=\sqrt{M_{ADM}^2-a^2}, \quad a=\frac{S}{M_{ADM}^2}.
\end{equation}
As in the previous example, the extreme limit $\sqrt{S}\to M_{ADM}$ of
Kerr metric
correspond  $m^p\to 0$. The explicit expression of these data can be found
in~\cite{Avila:2008te}, in this reference it is proved that the conformal
factor behaves in a similar way as the  Reissner-Nordstr\"om in the
limit $m^p\to 0$. For completeness, we reproduce this calculation and
adapt it to our setting.  
We use the coordinate
transformation~\cite{Brandt:1996si}
\begin{equation}
  r = \bar r \left(1+\frac{m+a}{2 \bar r}\right)
    \left(1+\frac{m-a}{2 \bar r}\right),
\end{equation}
where $r$ is the standard Boyer-Lindquist radial coordinate, which
puts the Kerr metric in the quasi-isotropic form
\begin{equation}
  ds^2 = \phi_K^4 \left [e^{-2 q_K}
   \left(d\bar r^2 + \bar r^2 d\theta^2\right) +
   \bar r^2 \sin\theta^2 d\phi^2\right],
\end{equation}
where $ds^2$ is the spatial line element,
\begin{equation}
e^{-2q_K}=\frac{r^2+a^2\cos^2\theta}{r^2+a^2+
    \frac{2ma^2r\sin^2\theta}{r^2+a^2\cos^2\theta}},
\end{equation}
and
\begin{equation}
 \phi^4_K =\bar{r}^{-2}\left[r^2+a^2+
    \frac{2ma^2r\sin^2\theta}{r^2+a^2\cos^2\theta}\right].
\end{equation}
The function $\phi^4$ has the expansion
\begin{eqnarray}
  \label{eq:KerrCF}
\phi_K^4 &=&  \frac{\left(a^2-m^2\right)^2}{16 \bar r^4}+
   \frac{m^3-a^2 m}{2 \bar r^3}+\frac{a^2+3 m^2}{2 \bar r^2}+ \\
   && \frac{2 m \left(-2 \cos (2 \theta )
   a^2+a^2+m^2\right)}{\left(m^2-a^2\right)
   \bar r}+
{\cal O}\left(1 \right)
\end{eqnarray}
when $a<m$. However, for $a=m$ expansion (\ref{eq:KerrCF}) is
singular. In this case $\phi^4$ has the form:
\begin{equation}
  \phi_K^4 = \frac{4 m (m+\bar r) \left(2 m^2+2 \bar r
   m+\bar r^2\right)}{\bar r^2 \left(\cos (2 \theta )
   m^2+3 m^2+4 \bar r m+2 \bar r^2\right)}+1,
\end{equation}
which has the expansion
\begin{equation}
 \phi_K^4 =\frac{8 m^2}{(\cos (2 \theta )+3) \bar r^2}+
   \frac{32 m \cos ^2(\theta )}{(\cos (2 \theta )+3)^2
   \bar r}+{\cal O}\left(1\right).
\end{equation}
Thus the behavior of $\phi_K$ changes from
$$
\phi_K\sim\sqrt{a^2-m^2}/(2 \bar r) +{\cal O}(1)
$$ 
to
$$
  \phi_K \sim\frac{\sqrt{2 m /\bar r}}{\sqrt[4]{1+\cos^2\theta}}
    + {\cal O}(r^{1/2})
$$ 
in the extremal case $a=m$
(the horizon is then mapped to the limiting surface $\bar r=0$).
Hence, just as in the maximal Bowen-York case, the $\bar r=0$
coordinate singularity corresponds to an end which is not 
asymptotically flat.

\subsection{Numerical Test of Highly-Spinning BY Initial Data}
\label{sec:hspinnrtest}

\begin{figure}
\begin{center}
\caption{The $\gamma_{xx}$ component of the metric on the initial
slice for a single spinning, non-boosted black hole located at the
origin with $S = 1M^2$ and the puncture mass parameter $m^p = 0$. Here
we plot $\gamma_{xx}$ versus $x$ ($y=z=.00175M$) and $z$
($x=y=0.00175M$) and fits to $\gamma_{xx} = a + b / (x^2 + 2 c^2)$
(with a similar form for $z$).  Note that $\gamma_{xx}\propto 1/r^2$
is consistent with the expected $\varphi \propto 1/\sqrt{r}$.  }
\includegraphics[width=3.3in]{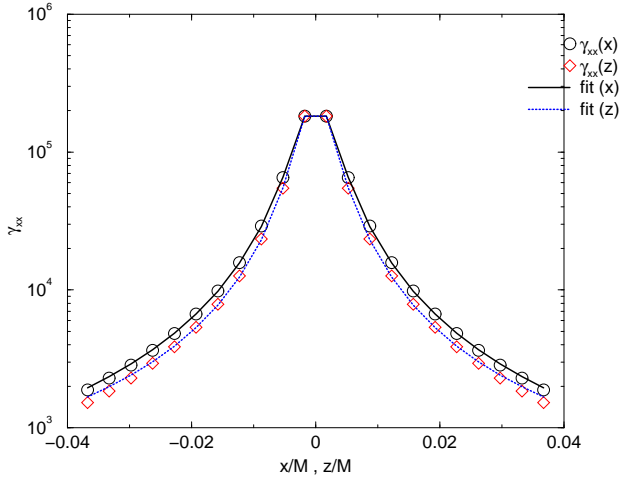}
\label{fig:gxx_max}
\end{center}
\end{figure}

\begin{figure}
\begin{center}
\caption{The function $u$ for finite $m_p$ in the neighborhood of the
puncture.
Here the data consists of two BH (the second BH is located at $x/M=-20$ 
and is not shown). The BH at the origin has spin parameter $S/M^2=1$.
Note that $u$ approaches $1/\sqrt{r}$ as $m_p$ tends to zero.}
\includegraphics[width=3.3in]{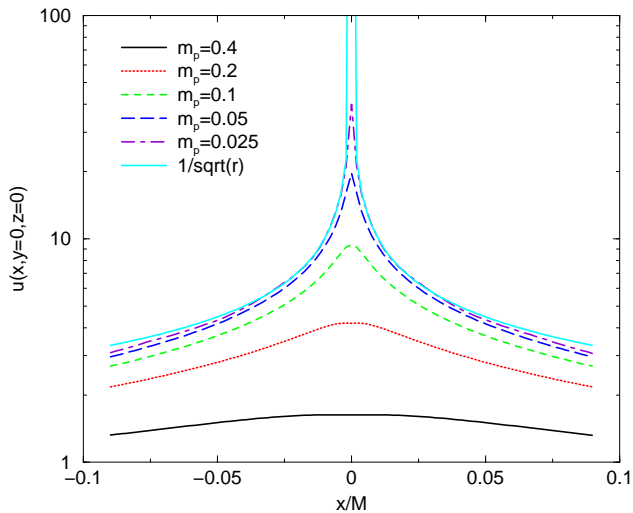}
\label{fig:u_v_mp}
\end{center}
\end{figure}

We first solve the initial data for one black hole presented in
section \ref{sec:BYID}. We used a modified version of BAM\_Elliptic
thorn~\cite{Brandt97b,cactus_web} in order
to solve for the case where $S/M^2=1$ (here $M$ denote an scale factor)
and $m^p = 0$ on a uniform grid
with resolution $h=0.0035 M$ and outer boundary at $0.64
M$ (here we are only interested in the singular behavior, which is not
affected by inaccurate boundary data). In order to
avoid the singularity itself, we constructed the grid such that the
origin was located halfway between gridpoints. We fit $\gamma_{xx}$ along
the lines $y=z=h/2$ and $x=y=h/2$ to the form:
$\gamma_{xx} = a + b / (x^2 + 2 c^2)$ (with a similar form
for $z$) and perform a non-linear
least-squares fit (See Fig.~\ref{fig:gxx_max}). We find that $\gamma_{xx}
(x) = 11.483 + 2.5042 / (x^2+ 2
(0.00231057)^2$ and $\gamma_{xx} (z) = 265.319 + 1.91325 / (x^2 + 2
(0.00192925)^2)$. Note that the $c$ parameter differs from the
expected $h/2$ but this functional form captures the expected singular behavior
to high accuracy. In Fig.~\ref{fig:u_v_mp} we  plot the function $u$
for various choices of mass parameter $m_p$ for a configuration
consisting of two black holes, one at the origin and one at $x=-20M$, 
with the
former hole having spin $S/M^2=1$ and the latter nonspinning. From the
plot one can see that, although $u$ is finite, $u$ ends to
$1/\sqrt{r}$ as $m_p$ tends to zero.

We are not able to calculate the ADM mass $M_{ADM}$ of the data,
because we needed very high resolution near the puncture $r=0$,
leaving too few point outside the horizon. Instead,
we compute the horizon mass $m_H$ given by 
${m_H} = \sqrt{m_{\rm irr}^2 + S^2/(4 m_{\rm irr}^2)}$
 where $m_{\rm irr}$ is the
irreducible mass (it is expected for these kind of data that $m_H\leq
M_{ADM}$, there exists however no proof of this conjecture).
Analogous to the ratio \eqref{eq:13} we define the quasi local
ratio $a/m_{H}$, where $a=S/m_{H}$.  For these data the maximum
possible value of this quantity is given by $a/m_{H} \sim
0.93$~\cite{Cook:1989fb, Dain:2002ee}.

Even the horizon mass $m_H$ is difficult to resolve at the initial
surface (the horizon is located at only one point $r=0$), we need to
perform the evolution of the data to compute it at later times.

\begin{figure}
\begin{center}
\caption{The horizon mass and specific spin for a maximal BY
black hole. Here $S=1M^2$ and $m^p=0$. Note that the spin drops
rapidly to about $a/m_{H} = 0.934$ (and is still dropping) after
the hole absorbs significant mass. The expected asymptotic value is
$a/m_{H} = 0.928$.
}
\includegraphics[width=3in]{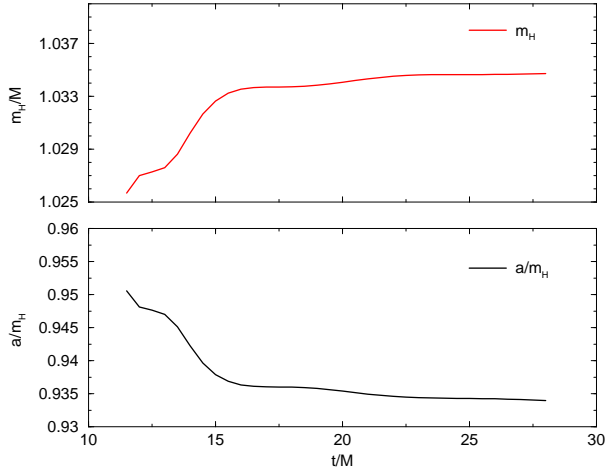}
\label{fig:max_early_hor}
\end{center}
\end{figure}

These data have axial symmetry and then the spin is a conserved
quantity. One part of the radiation emitted by the data will fall into
the black hole increasing its area and the other part will scape to
infinity.  Hence the ratio $a/m_{H}$ will monotonically decrease
during the evolution. This is precisely what we observed in the
numerical evolution presented in Fig. \ref{fig:max_early_hor}.

In order to measure the horizon mass and the quotient $a/m_{H}$ in
this case, we performed a second unigrid run, this time using
fisheye~\cite{Baker:2001sf,Campanelli:2006gf}, with a central
resolution of $h=M/64$ and outer boundaries at $65M$. Initially the
horizon has a coordinate radius of zero (akin to extreme Kerr in
quasi-isotropic coordinates) that grows to $r\sim 0.3M$ at $t=15M$ as it
absorbs the spurious radiation produced by the conformally flat
initial data.  In Fig.~\ref{fig:max_early_hor} we show the horizon
mass and specific quotient for this run. Note the rapid drop off in
the ratio $a/m_{H}$ between $t=10M$ and $t=15$. Due to the
relatively poor resolution of this run, and the short evolution time,
it is not clear if it asymptotes to $a/m_{H} = 0.933$, or will continue
to drop to the numerically predicted value of $0.928$.

\subsection{Spinning-black-hole-binary initial data}\label{sec:ID}

The initial data techniques of Sec.~\ref{sec:BYID} can be extended to
multiple spinning black holes with linear momentum~\cite{Brandt97b}.
Here too the 3-metric on the initial slice has the form $\gamma_{a b}
= (\psi_{BL} + u)^4 \delta_{a b}$, where $\psi_{BL}$ is the
Brill-Lindquist conformal factor, $\delta_{ab}$ is the Euclidean
metric, and $u$ is (at least) $C^2$ on the punctures.  The
Brill-Lindquist conformal factor is given by $ \psi_{BL} = 1 +
\sum_{i=1} m_{[i]}^p / (2 |\vec r - \vec r_{[i]}|),$ where the sum is
over all punctures,  $m_{[i]}^p$ is the mass parameter of puncture $i$
($m_{[i]}^p$ is {\em not} the horizon mass associated with puncture
$i$), and $\vec r_{[i]}$ is the coordinate location of puncture $i$
(we use the notation $[i]$ to distinguish the puncture label from the
tensor indices in the equations below). The
extrinsic curvature $K_{ab}$ is given by the Bowen-York (BY)
ansatz~\cite{Bowen80} and has the form $K^{a b} = \varphi^{-10} \tilde
K^{ab}$, where
\begin{eqnarray}
  \tilde K^{a b} = \sum_i &&\frac{3}{2 |\vec r - \vec r_{[i]}|^2} (2 P_{[i]}^{(a}
n_{[i]}^{b)} -
(\delta^{a b} - n_{[i]}^a n_{[i]}^b) P_i^c n_{[i]c}) \nonumber \\
&+&\frac{3}{|\vec r - \vec r_{[i]}|^3} (2 S_{[i]c} n_{[i]d}\epsilon^{c d (a}
n_{[i]}^{b)}),
\label{eq:BYK}
\end{eqnarray}
$\varphi = \psi_{BL} + u$,
$\vec n_{[i]} = (\vec r - \vec r_{[i]}) / |\vec r - \vec r_{[i]}|$,
and $\vec P_{[i]}$ and $\vec S_{[i]}$ are the linear and angular
momenta of puncture $i$.  As shown in Sec.~\ref{sec:BYID}, for a
single puncture, if $m^p =0$, $\vec P=0$, and $\vec S \neq 0$ then $u$
is no longer finite at the puncture location, but has a $1/\sqrt{r}$
singularity. Under these conditions, the resulting black hole will
have the maximum possible specific spin for Bowen-York type data.

\begin{figure}
\includegraphics[width=3.4in]{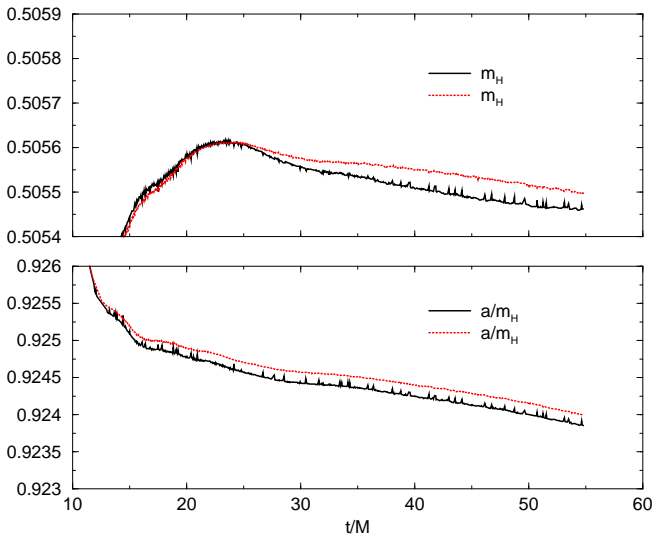}
\caption{The specific spins and horizon masses for a binary containing
nearly maximal BY spinning holes. The initial spins of the two
holes are $a/m_{H} = 0.967$.
The individual horizon masses increase rapidly near
$t\sim15M$ as the black holes absorb the spurious radiation. The 
spin drops to $a/m_{H} = 0.924$ since the spurious radiation does
not increase the angular momentum of the holes.}
\label{fig:s92_early_horizons}
\end{figure}

We compute and evolve the black-hole-binary data described above. 
The ratio $a/m_H$ is a quasi local quantity
which is  well defined for individual black holes in a binary system
(in contrast with the ratio $\epsilon_S$ in which appears the ADM mass
which is a global quantity).  We can obtain specific ratios $a/m_H$
nearly equal to the maximum allowed for BY holes by setting $m^p$
sufficiently small.  For the runs presented below we use the puncture
approach~\cite{Brandt97b} along with the {\sc
  TwoPunctures}~\cite{Ansorg:2004ds} thorn to compute initial data for
black-hole binaries.  In Fig.~\ref{fig:s92_early_horizons} we show the
isolated horizon spin for the individual holes in a near maximally
spinning BY binary (configuration MR0 described below).  Note that the
spins drop significantly within $10M$ -- $20M$ of evolution.

\section{Numerical Techniques}\label{sec:Techniques}

We evolve the black-hole-binary data-sets
 using the {\sc
LazEv}~\cite{Zlochower:2005bj,Campanelli:2005dd} implementation of the 
moving puncture approach~\cite{Campanelli:2005dd,Baker:2005vv}.  In our version of the moving
puncture approach we replace the
BSSN~\cite{Nakamura87,Shibata95, Baumgarte99} conformal exponent
$\phi$, which has logarithmic singularities at the punctures, with the
initially $C^4$ field $\chi = \exp(-4\phi)$.  This new variable, along
with the other BSSN variables, will remain finite provided that one
uses a suitable choice for the gauge. An alternative approach uses
standard finite differencing of $\phi$~\cite{Baker:2005vv}.

We use the Carpet~\cite{Schnetter-etal-03b} mesh refinement
driver to provide a `moving boxes' style mesh refinement. In this
approach  refined grids of fixed size are arranged about the
coordinate centers of both holes.  The Carpet code then moves these
fine grids about the computational domain by following the
trajectories of the two black holes.

We obtain accurate, convergent waveforms and horizon parameters by
evolving this system in conjunction with a modified 1+log lapse and a
modified Gamma-driver shift
condition~\cite{Alcubierre02a,Campanelli:2005dd}, and an initial lapse
$\alpha(t=0) = 2/(1+\psi_{BL}^{4})$.
The lapse and shift are evolved with
\begin{subequations}
\label{eq:gauge}
\begin{eqnarray}
(\partial_t - \beta^i \partial_i) \alpha &=& - 2 \alpha K\\
 \partial_t \beta^a &=& B^a \\
 \partial_t B^a &=& 3/4 \partial_t \tilde \Gamma^a - \eta B^a.
 \label{eq:Bdot}
\end{eqnarray}
\end{subequations}
These gauge conditions require careful treatment of $\chi$, the
inverse
of the three-metric conformal factor,
near the puncture in order for the system to remain
stable~\cite{Campanelli:2005dd,Campanelli:2006gf,Bruegmann:2006at}. In
Ref.~\cite{Gundlach:2006tw} it was
shown that this choice of gauge leads to a strongly hyperbolic
evolution system provided that the shift does not become too large.

We use {\sc AHFinderdirect}~\cite{Thornburg2003:AH-finding} to locate
apparent horizons.
We measure the magnitude of the horizon spin using the Isolated
Horizon~\cite{Ashtekar:2000hw,Dreyer:2002mx,
Schnetter:2006yt,Campanelli:2006fg,Campanelli:2006fy,Krishnan:2007pu}
algorithm detailed in~\cite{Dreyer02a}. This algorithm is
based on finding an approximate rotational Killing vector (i.e.\ an
approximate rotational symmetry) on the horizon, and given this
approximate Killing vector $\xi^a$, the spin magnitude is
\begin{equation}\label{isolatedspin}
S_{[\xi]} = \frac{1}{8\pi}\oint_{AH}(\xi^aR^bK_{ab})d^2V
\end{equation}
where $K_{ab}$ is the extrinsic curvature of the 3D-slice, $d^2V$ is
the
natural volume element intrinsic to the horizon, and $R^a$ is the
outward pointing unit vector normal to the horizon on the 3D-slice.
We measure the
direction of the spin by finding the coordinate line joining the poles
of this Killing vector field using the technique introduced
in~\cite{Campanelli:2006fy}.  Our algorithm for finding the poles of
the Killing vector field has an accuracy of $\sim 2^\circ$
(see~\cite{Campanelli:2006fy} for details).

We measure radiated energy, linear momentum, and angular momentum, in
terms of $\psi_4$, using the formulae provided in
Refs.~\cite{Campanelli99,Lousto:2007mh}. However, rather than using
the full $\psi_4$ we decompose it into $\ell$ and $m$ modes and solve
for the radiated linear momentum, dropping terms with $\ell \geq 5$.
The
formulae in Refs.~\cite{Campanelli99,Lousto:2007mh} are valid at
$r=\infty$. We obtain highly
accurate values for these quantities by solving for them on spheres of
finite radius (typically $r/M=25, 30, 35, 40$), fitting the results to
a polynomial dependence in $l=1/r$, and extrapolating to $l=0$. We
perform fits based on a linear and quadratic dependence on $l$, and
take the final values to be the average of these two extrapolations
with the differences being the extrapolation error.

\section{Results}\label{sec:results}
Evolving black holes with specific spins of $a/m_{H}\sim0.92$ 
is challenging because the
horizons appear quite small. In our coordinates, the initial horizon radii
were
$0.04M$. We evolved these data using 14 levels of refinement
with a finest resolution of $h=M/320$ (the highest resolution
reported so far in numerical simulations of binary black holes). 
The outer boundaries were
located at $\pm640M$ in all directions and the resolution on the
coarsest grid was $h=12.8M$. We obtained the spin, position, and
momentum parameters of the initial data using third-order post-Newtonian
parameters for equal-mass quasi-circular binaries with spins aligned
with the linear momentum of the two holes, and aligned with the
orbital angular momentum. In all cases we took the spin of the two
holes to be $a/m_{H} = 0.92$. We set the puncture mass parameters 
of the two holes such that the total ADM mass was 1M. The initial data
parameters are summarized in Table~\ref{table:BYparams}. Note that we
normalize the MR0 configuration to ADM mass of 1M, but keep the same
mass parameters when we rotate the spin. Hence MR45 -- MR315 have
slightly different ADM masses.
\begin{table}
\caption{Initial data parameters for the maximum recoil configurations
MR0 -- MR315 and the maximum hangup configuration MH.  For MR0 -- MR315
the punctures are located at $\vec x/M = (\pm 3.564036838,0,0)$, with
momenta $\vec p/M = (0,\pm 0.1254868859 ,0)$ and spins $\vec S =\pm(S_x, S_y,
0)$.  In all cases the orbital frequency is $M\omega = 0.045$ and the
puncture mass parameters are $m^p/M = 0.08967$. For MH the punctures
are located at $\vec X/M = (\pm 4.083304018,0,0)$, with momenta $\vec
P = (0,\pm 0.1046285561 ,0)$ and spin $\vec S/M^2 =
+(0,0,0.2362330001)$.  The orbital frequency is $M\omega = 0.035$
and the puncture mass parameters are $m^p/M = 0.107949$.  }
\begin{ruledtabular}
\begin{tabular}{lccc}
Config & $M_{\rm ADM}$ & $S_x/M^2$ & $S_y/M^2$ \\
MR0   & 1.000000 & $0$ & $0.23642497$ \\
MR45  & 0.999493 & $-0.16717770$ & $0.16717770$ \\
MR90  & 0.998982 & $-0.23642497$ & $0$ \\
MR135 & 0.999493 & $-0.16717770$ & $-0.16717770$ \\
MR180 & 1.000000 & $0$ & $-0.23642497$ \\
MR225 & 0.999493 & $0.16717770$ & $-0.16717770$ \\
MR270 & 0.998982 & $0.23642497$ & $0$ \\
MR315 & 0.999493 & $0.16717770$ & $0.16717770$ \\
\hline
MH    & 1.000000 \\
\end{tabular} \label{table:BYparams}
\end{ruledtabular}
\end{table}

\subsection{Large Recoil Velocities}
In Ref.~\cite{Campanelli:2007ew} we proposed a semi-empirical formulae
for the dependence of the merger recoil velocity on the spins and mass
ratio of the two black holes in a binary. Our formula predicts that
the largest recoils occur for two equal-mass, equal-spin black holes
with spins pointing in the orbital plane and counter aligned with each
other. In this configuration the recoil is proportional to the
spin-amplitude and varies sinusoidally with the angle between the direction
of the spins at merger and the linear momentum direction. The
resulting recoil will be perpendicular to the orbital plane. We were
able to test this prediction~\cite{Campanelli:2007cga} by evolving a binary 
with spins
$a/m_{H}=0.5$ (initially pointing in the direction of the linear
momentum) and then evolving a set of binary configuration with the
same orbital parameter, but with the initial spin directions rotated
by an angle
$\Theta$. We found that we
could fit the resulting recoil velocities to a $\cos(\Theta -
\Theta_0)$
dependence to high accuracy. Implicit in this approach is that the
angle between the spins  and linear momentum at merger is given by the
angle at merger for the $\Theta=0$ configuration plus $\Theta$.
This, in turn, requires that the $xy$ projection of the trajectories
be independent of $\Theta$.

\begin{figure}
\caption{The $xy$ projections of the puncture trajectories for MR0 and
MR45, the latter rotated by an angle $\Theta_{\rm cor}$ so that the
late inspiral and merger phases overlap.}
\includegraphics[width=2.7in,height=2.52in]{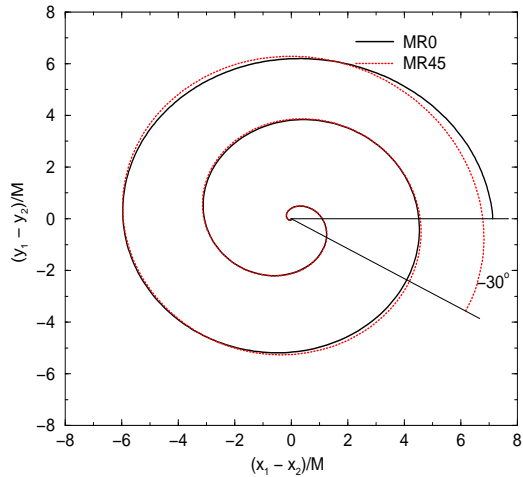}
\label{fig:0_45_track_rotate}
\end{figure}

\begin{figure}
\caption{Recoil velocity versus corrected rotation angle for the
MR0---MR315 configurations and a non-linear least-squares fit to a
simple sinusoidal behavior. Note that the corrected rotation angles
are not distributed uniformly in the range $(0,2 \pi)$.}
\includegraphics[width=3.4in]{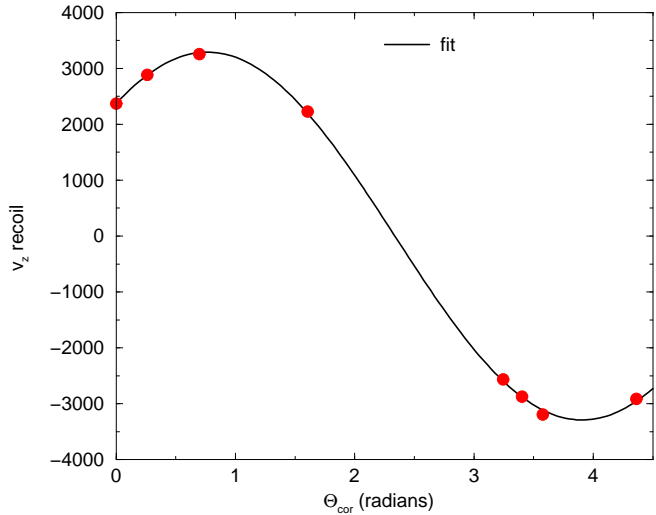}
\label{fig:kick_res}
\end{figure}

Unlike in Ref.~\cite{Campanelli:2007cga}, here we found for the MR0 ---
MR315 configurations that rotating the
initial spin direction
alters the $xy$-projection of the trajectories by introducing varying
ellipticity to the orbit and possibly due to spin-orbit coupling.
However, after rotating the $xy$-projected trajectories for M45 --
M315 by an angle $\Theta_{\rm rot}$ (See
Fig.~\ref{fig:0_45_track_rotate}, and Ref.~\cite{Lousto:2007db} for 
a description of the technique),
 we found that
they overlap for the late inspiral and plunge with projected MR0
trajectory. 
We then take the initial spin orientation, plus this trajectory
rotation, as the angle between MR45 -- MR315 spin direction at merger
with the MR0 spin direction at merger. We then fit the $z$-component
of the recoil versus this angle. The results are summarized in
Table~\ref{table:kick_res} and Fig.~\ref{fig:kick_res} (with several
runs having significantly higher  recoil velocities than any previous
simulation). A fit
of $v_z$ versus $\Theta_{\rm cor}$ (where $\Theta_{\rm cor}$ is the 
corrected angle between MR45 -- MR315) and MR0 gives:
$v_z (\KMS) = 3290.14 \cos (\Theta_{\rm cor} - 0.765885)$
where $\Theta_{\rm cor}$ is measured in radians and the confidence
interval for the amplitude is $(3243, 3337)$. Our empirical formula
predicts an amplitude of $3461\pm58$. These results are with 2.3
$\sigma$ of the prediction for $a/m_{H}=0.923$. A relatively large
error is not unexpected due to the difficulty in evolving systems with
such small scale features and the ellipticity introduced by the spurious
radiation. On the other hand, this may also indicate a non-linear (in
$a/m_{H}$) term is present that reduces the maximum possible recoil.
\begin{table}
\caption{Recoil velocity (in the $z$-direction) for the MR0---MR315
configuration, initial angle $\Theta_{\rm con}$ between the
spin directions of the MR45---MR315 and MR0, and
(approximate) angle $\Theta_{\rm cor}$ between spin directions of
MR45---MR315 and MR0 at merger.
Here $\Theta_{\rm cor} = \Theta_{\rm con}  + \Theta_{\rm rot}$.}
\begin{ruledtabular}
\begin{tabular}{lccc}
Config & $\Theta_{\rm con}$ & $\Theta_{\rm cor}$ & $V_z (\KMS)$\\
MR0   & $0^\circ$ & $0^\circ$ & $2372\pm12$\\
MR45  & $45^\circ$ & $15^\circ$ & $2887\pm27$\\
MR90  & $90^\circ$ & $40^\circ$ & $3254\pm19$\\
MR135 & $135^\circ$ & $92^\circ$ & $2226\pm6$\\
MR180 & $180^\circ$ & $186^\circ$ & $-2563\pm8$\\
MR225 & $225^\circ$ & $195^\circ$ & $-2873\pm23$\\
MR270 & $270^\circ$ & $205^\circ$ & $-3193\pm45$\\
MR315 & $315^\circ$ & $250^\circ$ & $-2910\pm1$\\
\hline
\end{tabular} \label{table:kick_res}
\end{ruledtabular}
\end{table}

\subsection{Orbital Hangup}

\begin{figure}
\caption{The puncture trajectories and first common apparent horizon
for the MH configuration. Note that the binary completes $\sim7.5$
orbits prior to merger.}
\includegraphics[width=3.0in,height=2.93in]{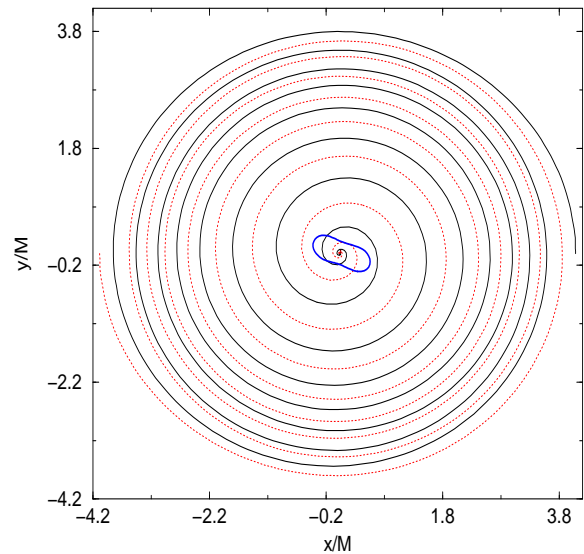}
\label{fig:mh_track}
\end{figure}

\begin{figure}
\caption{The puncture separation, and first derivative (see inset),
 versus orbital phase for the
MH configuration. Note the decaying oscillations that indicate the
ellipticity is significantly reduced after 2 orbits.}
\includegraphics[width=3.0in]{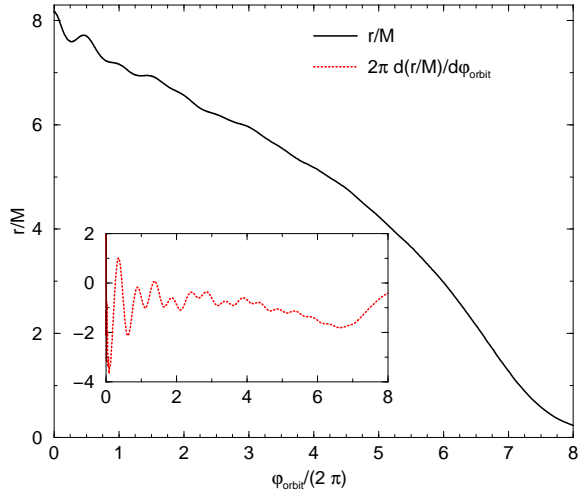}
\label{fig:mh_track_rvth}
\end{figure}

\begin{figure}
\caption{The $xy$ projection of the trajectory difference $\vec r_1 -
\vec r_2$ for the MH and MR0 configurations. Note the significant
hangup effect when the spins are parallel to the orbital angular
momentum (MH).}
\includegraphics[width=3.0in,height=3.0in]{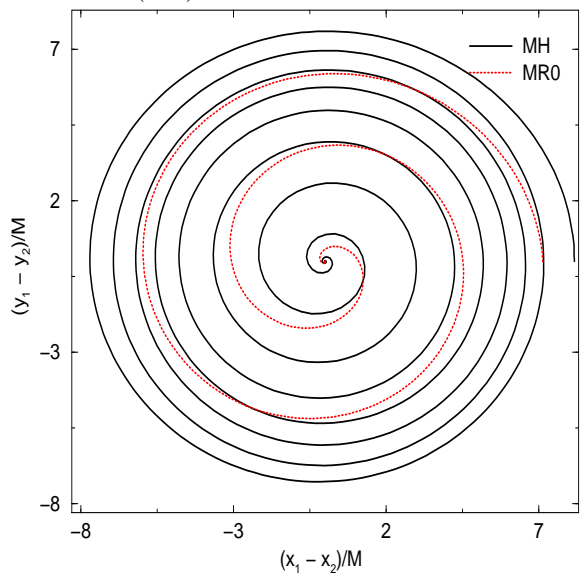}
\label{fig:mh_mr0_track}
\end{figure}

\begin{figure}
\caption{The real part of the $(\ell=2,m=2)$ mode of $\psi_4$ for 
the MH configuration showing the orbital dynamics and quasi-normal
decay.}
\includegraphics[width=3.4in]{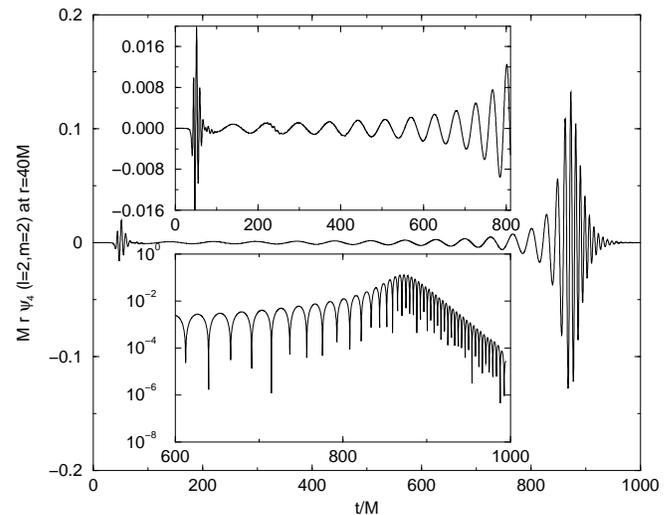}
\label{fig:mh_wave}
\end{figure}

  Of particular interest is the spin-orbit hangup
effect~\cite{Campanelli:2006uy,Hannam:2007wf,Rezzolla:2007rd} when
the two spins are aligned with the orbital angular momentum. Here we
examine configuration MH,  where the two spins have near maximal
(for BY data) spins. We evolved configuration MH with 14 levels of
refinement, maximum resolution of $h=M/320$, and outer boundaries at
$1281M$. In Fig.~\ref{fig:mh_track} we show the $xy$ trajectories of the
punctures as well as the first common apparent horizon. Note that the
binary completes $\sim 7.5$ orbits before the first common apparent
horizon forms. In Fig.~\ref{fig:mh_track_rvth} we show $r=|\vec r_1 -
\vec r_2|$ (where $\vec r_i$ is the location of puncture $i$)
versus orbital phase $\phi_{\rm orbit}$. The initial eccentricity, as
 is evident by the oscillation in $r(\phi_{\rm orbit})$, damp with
time. The hangup effect is clearly seen in
Fig.~\ref{fig:mh_mr0_track}, which shows the $xy$ projections of the 
trajectory difference $\vec r_1 - \vec r_2$ for the MH and MR0
configurations.
In Fig.~\ref{fig:mh_wave} we show the real part of the $(\ell=2,m=2)$
mode of $\psi_4$ for the MH configuration.
We measure the remnant mass and spin three different ways: from the
isolated horizon formalism, from the radiated energy and angular
momentum, and from the quasi-normal decay of the late-time waveform.
Results for the isolated horizon calculation were affected by late-time
boundary effects due to relatively poor resolution in the outer
zones. The results are summarized in Table~\ref{table:mh_res}.
To calculate the horizon mass and spin from the quasi-normal modes we
used the results of~\cite{Echeverria89} and a fit to $\exp(-t/\tau)
\sin(\omega t - \theta_0)$ for the real and imaginary parts of the
$(\ell=2,m=2)$ mode of $\psi_4$ at $r=40M$. We found $\tau/M =
15.097\pm.0074$ and $M \omega = 0.76579\pm0.00006$ (the errors
reported are the differences in $\tau$ and $\omega$ for the real and
imaginary parts of $\psi_4$). We also calculated the remnant mass and
spin from the radiated energy and angular momentum. Here too,
inaccuracies due to relatively poor resolution in the outer zones
affect the calculation. For all three methods, the final spin is in
qualitative agreement with the prediction of $a/m_{H} = 0.928$
in~\cite{Campanelli:2006uy,Campanelli:2006fg}, but differs from the
prediction $a/m_{H} = 0.9400\pm0.0019$ in~\cite{Rezzolla:2007xa}
for initial specific spins of $0.92$.
\begin{table}
\caption{Remnant horizon mass and spin for configuration MH based on
the isolated horizon (IH) calculation, the radiated energy and angular
momentum, and the quasi-normal frequency (QNM).}
\begin{ruledtabular}
\begin{tabular}{lccc}
 & IH & Radiation & QNM \\
\hline
$m_{H}$   & $0.9095\pm 0.0005$ & $0.9162\pm0.0024$ & $0.9146\pm0.0002$ \\
$a/m_{H}$ & $0.922\pm 0.001$ & $0.928 \pm 0.015$ & $0.922\pm0.001$\\
\end{tabular} \label{table:mh_res}
\end{ruledtabular}
\end{table}

\section{Conclusion}\label{sec:conclusion}

We have evolved equal-mass, equal-spin black-hole-binary configuration
with nearly maximal BY spin of $a/m_{H} = 0.923$, the highest spins
simulated thus far, both for spins
pointing in the same direction as the orbital angular momentum and for
spins (counter-aligned) pointing in the orbital plane. In the former
case we see a significant orbital hangup (for seven orbits prior to
merger), and confirmed that the remnant spin is non-maximal and agrees
with our previous predictions based on a least-squares fit of remnant
spin versus initial spin~\cite{Campanelli:2006uy, Campanelli:2006fg,
Campanelli:2006fy,Baker:2003ds}.  While in latter case we find that the maximum
recoil is $v_{\rm recoil} = 3290\pm47\ \KMS$, in qualitative agreement
with our prediction of $3461\pm58\ \KMS$.  The deviation of our
measured recoil velocity from the predicted value is likely due to the
relatively poor effective resolution (i.e.\ the number of gridpoints
across the initial horizons), as well as eccentricities introduced by
the significant amount of spurious radiation or nonlinear corrections
to the kicks formula. We have confirmed the sinusoidal dependence of
the recoil on the initial spin direction and that the recoil varies
essentially
linearly with the magnitude of the spin for a fixed initial spin
direction.  Note that the measured recoil of $v_{\rm recoil} =
3254\pm19\ \KMS$ for the MR90 configuration is the largest recoil
velocity  measured for an actual simulation to-date, and would expel the
BH remnant from any known galaxy.

All our numerical calculations confirm cosmic censorship.  In
particular, highly spinning black holes, which are close to the
extreme Kerr limit, do not develop naked singularities. Moreover, the
system always decays asymptotically in time to a final black hole which
satisfy the inequality between mass and spin of the Kerr black hole.

There is a curious analogous behavior between the conformal factor of
the extreme spinning BY initial data as $r\to0$, i.e.
$\varphi\sim1/\sqrt{r}$, and the late-time behavior of the determinant
of the 3-metric for a Schwarzschild black hole~\cite{Hannam:2006vv,
Baumgarte:2007ht, Brown:2007tb, Hannam:2008sg} with the standard
moving-puncture choice for the gauge Eqs. (\ref{eq:gauge}). Here $r=0$
corresponds to the horizon on the initial slice, which has finite
surface area, while in the case of Schwarzschild, if we take the limit
$t\to\infty$ and then $r\to0$, we approach a sphere of finite surface
area inside the horizon. It seems that, as we increase the spin of the
black hole to its maximum allowed value, we find a new, extreme
solution that has a different behavior from the non-extreme cases
($1/r$ vs. $1/\sqrt{r}$).  The limit of stationarity, for $t\to\infty$
of sub-maximal data, also leads to a new behavior for the conformal
factor $\phi\sim1/\sqrt{r}$, not present at finite time. In both cases
the data transition from a slicing that contains two asymptotically
flat ends to  one that contains one asymptotically flat end and one
cylindrical end. 

\begin{acknowledgments} 
We thank Manuela Campanelli for valuable discussions.  We gratefully
acknowledge NSF for financial support from grant PHY-0722315,
PHY-0653303, PHY 0714388, and PHY 0722703; and NASA for financial
support from grant NASA 07-ATFP07-0158.  Computational resources were
provided by the NewHorizons cluster at RIT and the Lonestar cluster at
TACC. Sergio Dain is supported by CONICET (Argentina).  This work was
supported in part by grant PIP 6354/05 of CONICET (Argentina),
Secyt-UNC (Argentina) and the Partner Group grant of the Max Planck
Institute for Gravitational Physics, Albert-Einstein-Institute
(Germany).

\end{acknowledgments} 

\bibliographystyle{apsrev}
\bibliography{../../../Lazarus/bibtex/references}

\end{document}